# Tunnel magnetoresistance of polymeric chains

Kamil Walczak [1]

Institute of Physics, Adam Mickiewicz University
Umultowska 85, 61-614 Poznań, Poland

Coherent spin-dependent electronic transport is investigated in a molecular junction made of polymeric chain attached to two ferromagnetic electrodes (Ni and Co, respectively). Molecular system is described by a simple Hückel model, while the coupling to the electrodes is treated through the use of a broad-band theory. The current flowing through such device is calculated within non-equilibrium Green's function approach. It is shown that tunnel magnetoresistance of molecular junction can be quite large (over 100 %) and strongly depends on: (i) the length of the polymeric chain and (ii) the strength of the molecule-to-electrodes coupling.

## I. Introduction

Development of the computer industry is conditioned by the miniaturization of circuit components. Molecular junctions are promising candidates as basic electronic devices because of their small size and self-assembly features. Such junctions are usually composed of two electrodes joined by organic molecule [1,2]. In general, transport properties of such structures are dominated by some effects of quantum origin, such as: (i) quantum tunneling, (ii) quantization of molecular energy levels and (iii) discreteness of electron charge. However, few years ago it was pointed out that also electron's spin as well as its charge can be employed to store, process and transmit information [3,4]. Spin-dependent transport is of great interest due to its potential applications in the future nanoelectronics.

Recent experiments on Ni nanocontacts disclosed magnetoresistance values up to 280 % at room temperature for a few-atom contact [5]. Similar effects are also expected for a molecular junctions, where single molecular wires are attached to ferromagnetic electrodes [6,7]. In this case, tunnel magnetoresistance (TMR) is associated with: (i) the asymmetry of the density of states (DOS) for two spin channels in the ferromagnetic materials and (ii) the quantum tunneling phenomenon. Generally, the tunneling probability of the electron flowing through the molecule depends on few factors: (i) the relative orientations of the electrode magnetizations which can be changed from the parallel (P) to antiparallel (AP) by applying an external magnetic field, (ii) the electronic structure of the molecular wire, (ii) the nature of the coupling between the molecule to the electrodes and (iv) the location of the Fermi level in relation to molecular energy levels.

The purpose of the present work is to study the coherent spin-dependent electronic transport of polymers (linear chain of one or more benzene rings) symmetrically coupled to a pair of identical ferromagnetic electrodes (Ni and Co) with the help of terminal atoms (see Fig.1). Since only delocalized π-electrons dictate the transport properties of analyzed structures, molecular system is described by Hückel model (π-electron approximation) [8,9], while the coupling to the electrodes is treated through the use of a broad-band theory [10].

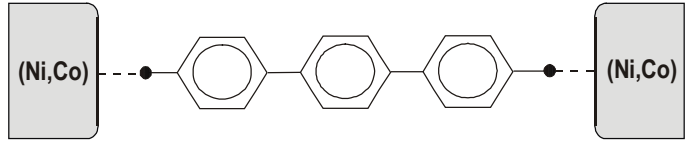

Fig.1 A schematic model of an ideal sample.

## II. Theoretical model

Hamiltonian for the system of two ferromagnetic electrodes joined by a polymeric chain is proposed in the following form:

$$H = \sum_{k,\sigma \in L,R} \varepsilon_{k,\sigma} c^+_{k,\sigma} c_{k,\sigma} + \sum_{\substack{k,\sigma \in L,R \\ m,\sigma \in M}} \left( t_{k\sigma,m\sigma} c^+_{k,\sigma} c_{m,\sigma} + h.c. \right) + H_M, \quad (1)$$

where first term describes electrons in the left (L) and right (R) electrodes, the second one is associated with the tunneling process between the electrodes and the molecule, while the third one corresponds to the tight-binding description of the molecular system (Hückel model):

$$H_M = \sum_{m,\sigma \in M} \varepsilon_{m,\sigma} c^+_{m,\sigma} c_{m,\sigma} + \sum_{m,\sigma \in M} \left( t_{m,m+1} c^+_{m,\sigma} c_{m+1,\sigma} + h.c. \right). \quad (2)$$

Here all the site energies ($\varepsilon$) and hopping integrals ($t$) are considered as model parameters.

The current flowing into the molecular system can be calculated from the time evolution of the occupation number $N_L = \sum_{k,\sigma \in L} c^+_{k,\sigma} c_{k,\sigma}$ for electrons in the left electrode [11-13]:

$$I = -e \frac{d}{dt} \langle N_L \rangle = \frac{e}{\hbar} \sum_{\substack{k,\sigma \in L \\ m,\sigma \in M}} t_{k\sigma,m\sigma} \int_{-\infty}^{+\infty} d\omega \left[ G^<_{m\sigma,k\sigma}(\omega) + c.c. \right], \quad (3)$$

where $G^<_{m\sigma,k\sigma}(\omega)$ is the Fourier transform of $G^<_{m\sigma,k\sigma}(t) \equiv i \langle c^+_{k,\sigma}(0) c_{m,\sigma} \rangle$. Using equation of motion technique to calculate the lesser Green's function, one can find an expression for the current given only by the bare Green functions of the electrodes and the dressed Green functions of the molecule [12]. Since the molecule is contacted with the electrodes only through the terminal atoms and densities of states for electrons in both electrodes are constant (by assumption), current formula can be rewritten in the form:

$$I = \frac{4e}{h} \sum_{\sigma \in L,R} \Delta_{L\sigma} \Delta_{R\sigma} \int_{-\infty}^{+\infty} d\omega \left[ f(\omega - \mu_L) - f(\omega - \mu_R) \right] | G_{1\sigma,N\sigma}(\omega) |^2, \quad (4)$$

where the summation is over two possible orientations of electron's spin ($\sigma = \uparrow, \downarrow$) and $G_{1\sigma,N\sigma}(\omega) = [\omega J - H_M - i\Delta_{L\sigma} - i\Delta_{R\sigma}]^{-1}_{1N}$ is matrix element taken between the first (1) and the last (N) terminal atoms in the wire (J denotes the unit matrix). In the above relation (Eq.4): $f(\omega - \mu_\alpha)$ denotes the equilibrium Fermi distribution function with electrochemical potential defined as $\mu_\alpha = \varepsilon_F \pm eV/2$ (for $\alpha = L$ and $R$, respectively), $\Delta_{\alpha\sigma} = \pi t^2_\alpha \rho_{\alpha\sigma}$, $t_\alpha$ is the hopping parameter responsible for the coupling with $\alpha$ electrode, and $\rho_{\alpha\sigma}$ is density of states at the Fermi energy level $\varepsilon_F$ for concrete spin $\sigma$ in the $\alpha$ electrode.

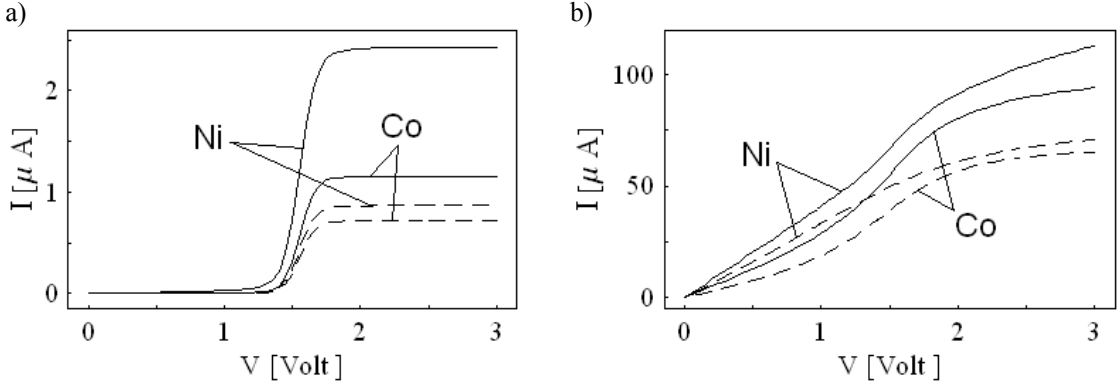

Fig.2 Current-voltage characteristics for the device made of one benzene ring attached to two ferromagnetic electrodes in the case of a) weak ($t_L = t_R = 0.05$ eV) and b) strong ($t_L = t_R = 0.5$ eV) coupling. Solid and broken lines correspond to parallel and antiparallel alignment of the electrodes' magnetization, respectively.

The voltage is assumed to be dropped entirely at the molecule/electrode interfaces so as the electronic structure of the molecule remains unaffected by the electrical field.

Tunnel magnetoresistance is defined as a relative change in the conductance of the system when the magnetizations of the two ferromagnetic layers switch between parallel (P) and antiparallel (AP) configurations, hence: $TMR = (G_P - G_{AP})/G_P$. Conductance itself is simply calculated as a derivative of the current (4) with respect to voltage, where: $G_P = G_{\uparrow\uparrow} + G_{\downarrow\downarrow}$ and $G_{AP} = G_{\uparrow\downarrow} + G_{\downarrow\uparrow}$ (the arrows denote the orientations of magnetization of the electrodes). Finally, it should be also mentioned that our calculations are based on the assumptions of coherent and elastic transport, for which the current conservation rule is fulfilled at each atom and for any energy $\omega$.

### III. Discussion of the results

In order to simulate conjugated molecules connected to ideal electrodes, we chose the following energy parameters (given in eV): $\varepsilon_m = 0$ (the reference energy), $t_{m,m+1} = -2.5$, $\varepsilon_F = 0$. Zero value of the Fermi energy corresponds to the assumption that this level is located exactly in the middle of the HOMO-LUMO gap of the isolated molecule.

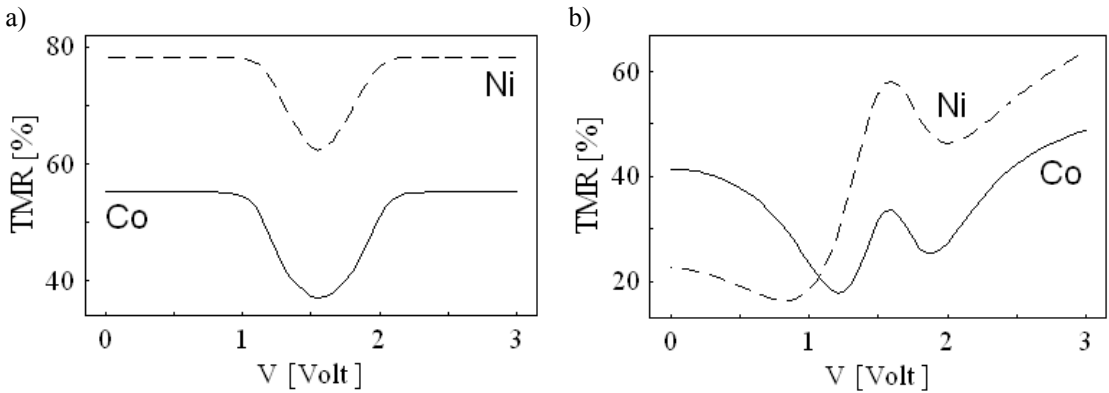

Fig.3 Tunnel magnetoresistance as a function of bias voltage for the device made of one benzene ring attached to two ferromagnetic electrodes in the case of a) weak ($t_L = t_R = 0.05$ eV) and b) strong ($t_L = t_R = 0.5$ eV) coupling. Solid and broken lines correspond to Co and Ni materials, respectively.

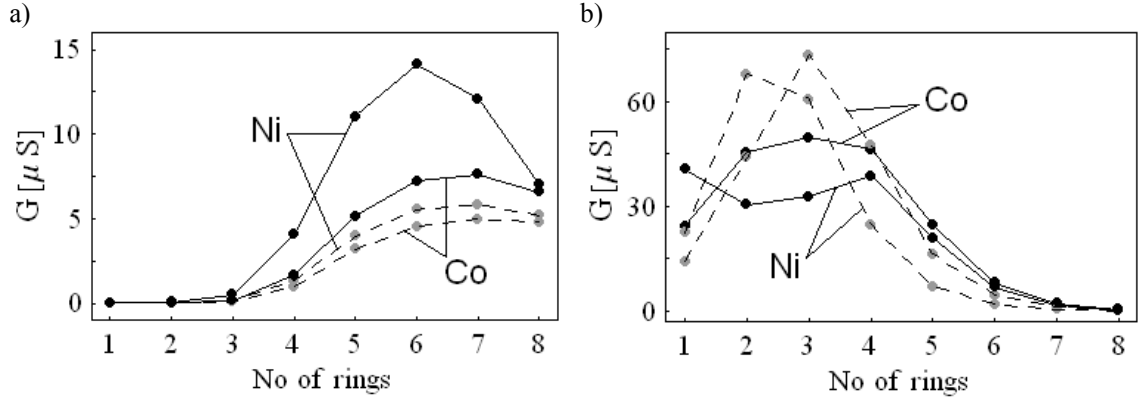

Fig.4 Conductance as a function of the number of benzene rings attached to ferromagnetic electrodes in the case of a) weak ($t_L = t_R = 0.05$ eV) and b) strong ($t_L = t_R = 0.5$ eV) coupling. Solid and broken lines (or black and gray circles) correspond to parallel and antiparallel alignment of the electrodes' magnetization, respectively.

Ferromagnets have unequal spin up and spin down populations, and therefore their densities of states for both spin orientations are different. The density of states for the Ni and Co electrodes are taken from the work [10], as obtained from the band structure calculations: $\rho_{Ni\uparrow} = 0.1897$, $\rho_{Ni\downarrow} = 1.7261$ and $\rho_{Co\uparrow} = 0.1740$, $\rho_{Co\downarrow} = 0.7349$ (given in states/eV). Since the polarization in the $\alpha$ electrode is defined as $P_\alpha = (\rho_{\alpha\uparrow} - \rho_{\alpha\downarrow})/(\rho_{\alpha\uparrow} + \rho_{\alpha\downarrow})$, therefore: $P_{Ni} = -0.802$ and $P_{Co} = -0.617$. Moreover, our calculations are performed at room temperature, but anyway all the results are not very sensitive to finite temperature (except the effect of thermal broadening).

Figure 2 shows the current-voltage (I-V) characteristics for one benzene ring attached to ferromagnetic electrodes. The jump in the I-V curve for the weak coupling case corresponds to the electron tunneling through the discrete energy level. The smoothening of the I-V dependence for the strong coupling case is associated with the energy level broadening due to its contact with the electrodes. The current for the P alignment reaches bigger values in comparison with the current for the AP configuration for particular materials (Ni and Co, respectively).

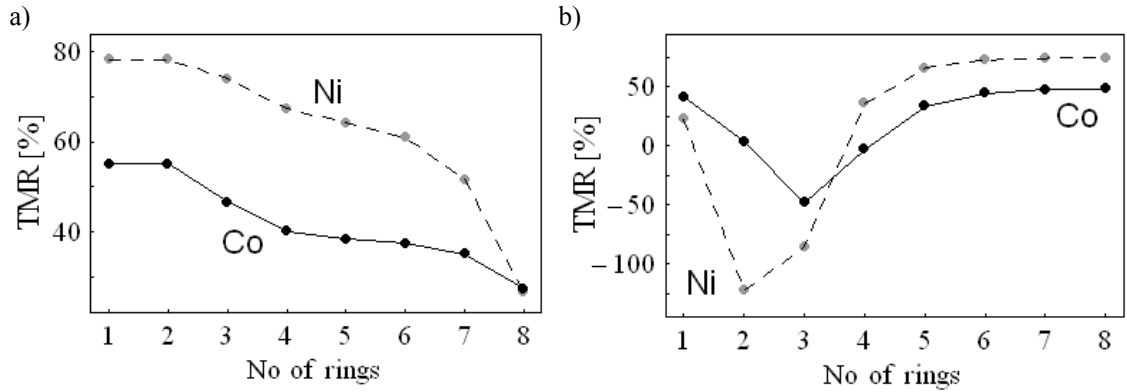

Fig.5 Tunnel magnetoresistance as a function of the number of benzene rings attached to ferromagnetic electrodes in the case of a) weak ($t_L = t_R = 0.05$ eV) and b) strong ($t_L = t_R = 0.5$ eV) coupling. Solid and broken lines (or black and gray circles) correspond to Co and Ni materials, respectively.

Usually the TMR coefficient increases with the increase of the polarization of the ferromagnetic electrodes, so we expect Ni to be more adequate than Co in generating the spin-valve effect. It is really true in the limit of weak coupling, where magnetoresistance takes the larger values of about 23 % for Ni in relation to Co electrode in the considered range of voltages (see Fig.3a). Moreover, for bias voltages far from the resonance region, TMR stabilizes at the constant value of about 78 % for Ni and 55 % for Co metals. However, in the vicinity of resonance we observe the valley in the magnetoresistance spectra, where TMR is reduced to 62 % for Ni and to 37 % for Co electrodes. Different behavior of TMR is predicted in the case of strong coupling. Namely, for $V < 1$ Volt, the TMR coefficient for Ni is smaller than for Co metal (see Fig.3b). Such result contradicts our intuition. Furthermore, at resonance we observe peak in the magnetoresistance spectra (instead of the valley found in the weak coupling limit).

For longer polymeric chains we can not ignore the voltage drop inside the molecule in the nonlinear response regime. However, we believe that at low voltages (small enough to treat the current as a linear function of bias), the method presented here is still effective in conductance calculations. In Fig.4 we show the changes of the conductance due to the increase of the number of benzene rings in the junction. In the weak coupling limit, the conductance values are always bigger for the P configuration than for the AP one (see Fig.4a). However, for the case of strong coupling, we can find few examples where conductance for the AP configuration is bigger than for the P one (2-3 rings for Ni electrode and 3-4 rings for Co electrode, as shown in Fig.4b). This behavior results in negative value of the TMR coefficient (check in Fig.5b). Although the transitions between positive and negative values of TMR with increasing the number of benzene rings are observed in the strong coupling limit, such transitions are unexpected for the case of the weak coupling (compare Figs.5a and 5b). Such unusual features of the TMR coefficients are due to the changes in the shape of molecular Green's function in the vicinity of the Fermi energy level with increasing the number of benzene rings in the junction [9].

## IV. A brief summary

Summarizing, we have performed the coherent spin-dependent calculations of transport characteristics for molecular junctions, which are composed of polymeric chain attached to two ferromagnetic electrodes (Ni and Co, respectively). All the results were obtained in the frames of non-equilibrium Green functions formalism, where molecular system were described by a simple Hückel model, while the coupling to the electrodes were treated through the use of a broad-band theory. Since analyzed junctions exhibit large values of magnetoresistance (sometimes it is much larger than for conventional TMR systems – even over 100 %), therefore such junctions are very interesting from the point of view of future applications. In particular, it is shown that the effect of spin-valve strongly depends on: (i) the length of the polymeric chain and (ii) the strength of the molecule-to-electrodes coupling. It should be also mentioned that the alignment of the electrode magnetizations can be controlled by applying an external magnetic field. However, the presence of magnetic field in the molecular region can have a direct influence on the shape of conductance [14].

**Acknowledgements**

The author is very grateful to B. Bułka for bringing to his attention the problem of spin-dependent transport in molecular structures and T. Kostyrko for valuable discussions. This work was supported in part by the State Committee for Scientific Research (Poland) within the project No. PBZ KBN 044 P03 2001.

**References**

[1] E-mail address: walczak@amu.edu.pl
[1] M. A. Reed, Proc. IEEE **87**, 625 (1999); and references therein.
[2] R. M. Metzger, Acc. Chem. Res. **32**, 950 (1999); and references therein.
[3] G. A. Prinz, Science **282**, 1660 (1998); and references therein.
[4] I. Žutić, J. Fabian, S. Das Sarma, Rev. Mod. Phys. **76**, 323 (2004).
[5] N. Garcia, M. Muñoz and Y.-W. Zhao, Phys. Rev. Lett. **82**, 2923 (1999).
[6] E. G. Emberly and G. Kirczenow, Chem. Phys. **281**, 311 (2002).
[7] M. Zwolak and M. Di Ventra, Appl. Phys. Lett. **81**, 925 (2002).
[8] T. Kostyrko, J. Phys.: Condens. Matter **14**, 4393 (2002).
[9] K. Walczak, Phys. Stat. Sol. (b) **241**, 2555 (2004).
[10] W. I. Babiaczyk and B. R. Bułka, Phys. Stat. Sol. (a) **196**, 169 (2003).
[11] Y. Meir and N. S. Wingreen, Phys. Rev. Lett. **68**, 2512 (1992).
[12] H. Huang and A.-P. Jauho, *Quantum Kinetics in Transport and Optics of Semiconductors* (Springer-Verlag, Berlin 1998).
[13] B. R. Bułka and S. Lipiński, Phys. Rev. B **67**, 024404 (2003).
[14] K. Walczak, Cent. Eur. J. Chem. **2**, 524 (2004).